\documentclass[aps,prd,10pt,twocolumn,nofootinbib,groupedaddress,superscriptaddress,floatfix]{revtex4-1}

\usepackage[colorlinks=true,urlcolor=blue,linkcolor=blue,citecolor=blue]{hyperref}
\usepackage{amsmath,amssymb,amstext,amsbsy,amsfonts,amsthm,graphicx,microtype,booktabs}
\usepackage[normalem]{ulem}
\usepackage[capitalize]{cleveref}

\newcommand{\abs}[1]{{\left \vert #1 \right \vert}}
\newcommand{\ud}{\textrm{d}}
\newcommand{\Mpl}{M_\mathrm{pl}}

\begin{document}

\title{Constraining early dark energy with gravitational waves before recombination}

\author{Zachary J. Weiner}
\email{zweiner2@illinois.edu}
\affiliation{Illinois Center for Advanced Studies of the Universe \& Department of Physics, University of Illinois at Urbana-Champaign, Urbana, Illinois 61801, USA}

\author{Peter Adshead}
\email{adshead@illinois.edu}
\affiliation{Illinois Center for Advanced Studies of the Universe \& Department of Physics, University of Illinois at Urbana-Champaign, Urbana, Illinois 61801, USA}

\author{John T. Giblin, Jr.}
\email{giblinj@kenyon.edu}
\affiliation{Department of Physics, Kenyon College, Gambier, Ohio 43022, USA}
\affiliation{CERCA/ISO, Department of Physics, Case Western Reserve University, Cleveland, Ohio 44106, USA}

\begin{abstract}
    We show that the nonperturbative decay of ultralight scalars into Abelian gauge bosons, recently
    proposed as a possible solution to the Hubble tension, produces a stochastic background of
    gravitational waves which is constrained by the cosmic microwave background.
    We simulate the full nonlinear dynamics of resonant dark photon production and the associated
    gravitational wave production, finding the signals to exceed constraints for the entire
    parameter space we consider.
    Our findings suggest that gravitational wave production from the decay of early dark energy may
    provide a unique probe of these models.
\end{abstract}

\maketitle


\section{Introduction}

Measurements of the current expansion rate $H_0$ as inferred from the acoustic peaks in the cosmic
microwave background (CMB) radiation are in tension with the value obtained from local
measurements~\cite{Riess:2019cxk,Aghanim:2018eyx,Freedman:2017yms}, suggesting the need to extend
the concordance $\Lambda$ cold dark matter cosmological model.
Rather than alter the expansion history at intermediate redshifts, new physics introduced to resolve
the tension must alter the absolute scales of the low- or high-redshift anchors of the cosmic
distance scale~\cite{Bernal:2016gxb, Aylor:2018drw}.
Early-Universe resolutions thus focus on altering the high-redshift anchor, the CMB sound horizon at
recombination.
For a recent review, see Ref.~\cite{Knox:2019rjx}.

In particular, an increased expansion rate before recombination decreases the sound horizon.
However, simply adding more radiation~\cite{Wyman:2013lza,Dvorkin:2014lea} also changes the damping
scale in a way that is increasingly disfavored by high-precision measurements of the high-multipole
damping tail~\cite{Hu:1996vq,Raveri:2017jto,Aghanim:2018eyx}.
To avoid this, one class of proposed solutions, so-called early dark energy (EDE) models, postulates
an additional energy component that is only transiently important near
recombination~\cite{Karwal:2016vyq, Poulin:2018dzj, Poulin:2018cxd, Lin:2019qug, Smith:2019ihp,
Berghaus:2019cls, Braglia:2020bym,Gonzalez:2020fdy,Niedermann:2019olb,Niedermann:2020dwg}.
These proposed solutions supposedly relieve the tension between the early and late data sets;
however, see Refs.~\cite{Krishnan:2020obg,Hill:2020osr}.

In the simplest EDE implementations, a scalar field is initially frozen up its potential in a
homogeneous configuration.
The field's mass is tuned such that it begins to evolve near matter-radiation equality, oscillating
about the minimum of its potential.
In order to redshift away fast enough (at least as fast as radiation), the potential must have no
quadratic term about its minima.
From a particle physics perspective, this requires an explanation; to avoid such extreme
fine-tuning, Ref.~\cite{Gonzalez:2020fdy} instead proposed a model of a \textit{decaying} ultralight
scalar (dULS).
Instead of oscillating about a peculiar potential and redshifting away, the EDE scalar field decays
resonantly to dark radiation during oscillations about a quadratic minimum.

Nonperturbative or resonant particle production is a common feature of early-Universe preheating
after inflation (for reviews, see Refs.~\cite{Allahverdi:2010xz,Amin:2014eta,Lozanov:2019jxc}).
Substantial study has established that these violent processes can also lead to the copious
production of gravitational waves~\cite{Khlebnikov:1997di,Easther:2006gt,Easther:2006vd,
Easther:2007vj,GarciaBellido:2007dg,Dufaux:2007pt,Dufaux:2010cf,Bethke:2013aba,Figueroa:2013vif,
Bethke:2013vca,Figueroa:2016ojl,Figueroa:2017vfa,Amin:2018xfe,Adshead:2018doq,Adshead:2019lbr,
Adshead:2019igv}.
The wavelength of the produced gravitational waves (GWs) is determined by the characteristic scale
of particle production, which must be smaller than the horizon size.
During preheating after inflation this restricts the production to frequencies from MHz to GHz, well
beyond the reach of current or planned detectors~\cite{TheLIGOScientific:2014jea,TheVirgo:2014hva,
Somiya:2011np,Audley:2017drz,Punturo:2010zz,Seto:2001qf,Corbin:2005ny}.\footnote{
    However, high-frequency gravitational waves contribute to the energy budget of the Universe as
    radiation.
    Constraints on the gravitational-wave energy density from
    $N_\mathrm{eff}$~\cite{Maggiore:1999vm,Pagano:2015hma,Lasky:2015lej,Meerburg:2015zua} can be used
    to indirectly restrict preheating
    scenarios~\cite{Adshead:2018doq,Adshead:2019lbr,Adshead:2019igv}.
}
By contrast, in order to resolve the Hubble tension, particle production due to the decay of the
ultralight scalar must occur near the time of matter-radiation equality.
Since resonant particle production occurs at scales near the horizon scale at that time,
gravitational wave emission occurs at current-day frequencies near $10^{-16} \, \mathrm{Hz}$.
CMB anisotropies constrain stochastic gravitational waves with peak sensitivity at present-day
frequencies near $10^{-17} \, \mathrm{Hz}$~\cite{Lasky:2015lej,Caprini:2018mtu, Clarke:2020bil,
Namikawa:2019tax}.
In this paper, we confront models of ultralight scalar decay into dark photons with these
constraints.


\section{Background}

Following Ref.~\cite{Gonzalez:2020fdy}, we study the resonant decay of an
ultralight axion $\phi$ into a (dark) Abelian gauge field $A_\mu$ described by the action
\begin{align}
    \begin{split}
    \label{eqn:action}
        S
        &= \int \ud^4 x \sqrt{-g} \Bigg[
            \frac{\Mpl^2}{2} R - \frac{1}{2} \partial_\mu \phi \partial^\mu \phi - V(\phi) \\
        &\hphantom{={} \int \ud^4 x \sqrt{-g} \Bigg[ }
            - \frac{1}{4} F_{\mu\nu} F^{\mu\nu}
            - \frac{\alpha}{4 f} \phi F_{\mu\nu}\tilde{F}^{\mu\nu}
            \Bigg].
    \end{split}
\end{align}
Here $f$ is the axion decay constant and $\alpha$ is a dimensionless coupling that parametrizes the
rate and efficiency of energy transfer.
The field-strength tensor of the dark photon is
$F_{\mu\nu} \equiv \partial_\mu A_{\nu} - \partial_\nu A_{\mu}$, while its dual is
$\tilde{F}^{\mu \nu} = \epsilon^{\mu \nu \alpha \beta} F_{\alpha \beta} / 2$,
where $\epsilon^{\mu\nu\rho\sigma}$ is the Levi-Civita symbol with convention
$\epsilon^{0123} = 1 / \sqrt{-g}$.
Following Ref.~\cite{Gonzalez:2020fdy}, we take the standard axion potential
$V(\phi) = m_\phi^2 f^2 (1 - \cos \phi / f)$.
We set $c = \hbar = k_B = 1$ and use $\Mpl = 1 / \sqrt{8 \pi G_N}$ to denote the reduced Planck
mass.
We work with the ``mostly plus,'' conformal Friedmann-Lema\^{\i}tre-Robertson-Walker (FLRW) metric,
$g_{\mu\nu} = a^2 \eta_{\mu\nu} = a^2 \, \mathrm{diag}\,(-1, 1, 1, 1)$, using primes to denote
derivatives with respect to conformal time $\tau$.

The classical equations of motion,
\begin{align}
    A_\nu''
    &= \partial_i \partial_i A_{\nu}
        + \eta_{\beta \nu} \frac{\alpha}{f} \partial_{\alpha} \phi
        \left(\frac{1}{2} \sqrt{-g} \epsilon^{\alpha \beta \rho \sigma} F_{\rho \sigma} \right)
        \label{eqn:gauge-eom-main-text} \\
	\phi''
    &= \partial_i \partial_i \phi
        - 2 \mathcal{H} \phi'
        - a^2 \frac{\ud V}{\ud \phi}
        - a^2 \frac{\alpha}{4 f} F_{\mu \nu} \tilde{F}^{\mu \nu},
    \label{eqn:phi-eom-main-text}
\end{align}
permit solutions in which fluctuations of the gauge fields are exponentially enhanced via a
tachyonic instability sourced by a homogeneous, rolling axion.
Initially, a cosmological axion has some static homogeneous component
$\left\langle \phi \right\rangle = \theta f$, expressed in terms of the initial misalignment angle
$\theta$.
As a result, the axion's energy is dominated by its potential, acting as a source of early dark
energy which could alleviate the Hubble tension~\cite{Poulin:2018cxd}.
On the other hand, to linear order the helical polarizations of $A_i$ obey~\cite{Adshead:2015pva}
\begin{align}
    {A_\pm}''(\mathbf{k})
        + k \left( k \mp \frac{\alpha}{f} \left\langle \phi' \right\rangle \right) A_\pm(\mathbf{k})
    = 0.
\end{align}
Thus, once the axion begins to oscillate (when the Hubble parameter drops below
$\sim m_\phi / 3$~\cite{Marsh:2015xka}), its non-negligible background velocity causes one of the
two polarizations to undergo tachyonic resonance, i.e., to be amplified exponentially for modes
$k < \alpha / f \times \left\langle \phi' \right\rangle$.

As the axion crosses zero, $\left\langle \phi' \right\rangle$ changes sign and so amplifies the
other polarization.
Eventually, the gauge field fluctuations become so large that nonlinear effects begin to fragment
the axion background, ending the phase of tachyonic resonance.
Both the initial exponential gauge field production and subsequent nonlinear dynamics can source
a significant gravitational wave background.
Gravitational waves correspond to the tensor part of perturbations to the spatial part of the
spacetime metric,
\begin{align}\label{eqn:hij-eom-main-text}
	h''_{ij} - \partial_k \partial_k h_{ij} + 2 \mathcal{H} h'_{ij}
    &= \frac{2}{\Mpl^2} T_{ij}^\mathrm{TT},
\end{align}
where $T_{ij}^\mathrm{TT}$ is the transverse and traceless part of the stress-energy tensor
$T_{ij}$.

We employ numerical simulations in order to fully capture resonance, the nonlinear dynamics which
terminate energy transfer, and the resulting production of gravitational waves.
We solve the classical equations of motion
\cref{eqn:gauge-eom-main-text,eqn:phi-eom-main-text,eqn:hij-eom-main-text} in a
homogeneous $\Lambda$CDM cosmology, self-consistently including the contribution of the dULS sector
to the expansion rate.
We discretize these equations onto a three-dimensional, periodic, regularly spaced grid, computing
spatial derivatives via fourth-order centered differencing and utilizing a fourth-order Runge-Kutta
method for time integration.
All results presented use grids with $N^3 = 768^3$ points, side length $L = 10 / m_\phi$, and a
time step $\Delta \tau = \Delta x / 10 = L / 10 N$.
We implement simulations using \textsf{pystella}~\cite{Adshead:2019lbr,Adshead:2019igv} and provide
details on our algorithm, initial conditions, and convergence tests in the Appendices.


\section{Results}

In our simulations of the decaying ultralight scalar model, we consider
benchmark scenarios from Ref.~\cite{Gonzalez:2020fdy}, taking
$m_\phi = 10^{-27} - 10^{-26} \, \mathrm{eV}$ so that the dULS sector transitions from dark energy
to matter-like behavior around the favored redshift $z_c \approx 16500$.
Because the axion begins to oscillate when $H \sim m_\phi$, its relative contribution to the
Universe's energy scales as $\rho_\phi / \rho \sim m_\phi^2 (\theta f)^2 / H^2 \sim (\theta f)^2$,
independent of $m_\phi$.
Thus, we set $f = 1.5 \times 10^{17} \, \mathrm{GeV}$ (and an initial misalignment angle $\theta =
2$ for convenience) so that the dULS sector makes up $\sim 3-4 \%$ of the Universe's energy budget
at its peak.
We comment later on the dependence of the gravitational wave signal on slight changes in these
choices.

\begin{figure}[t]
    \centering
    \includegraphics[width=\columnwidth]{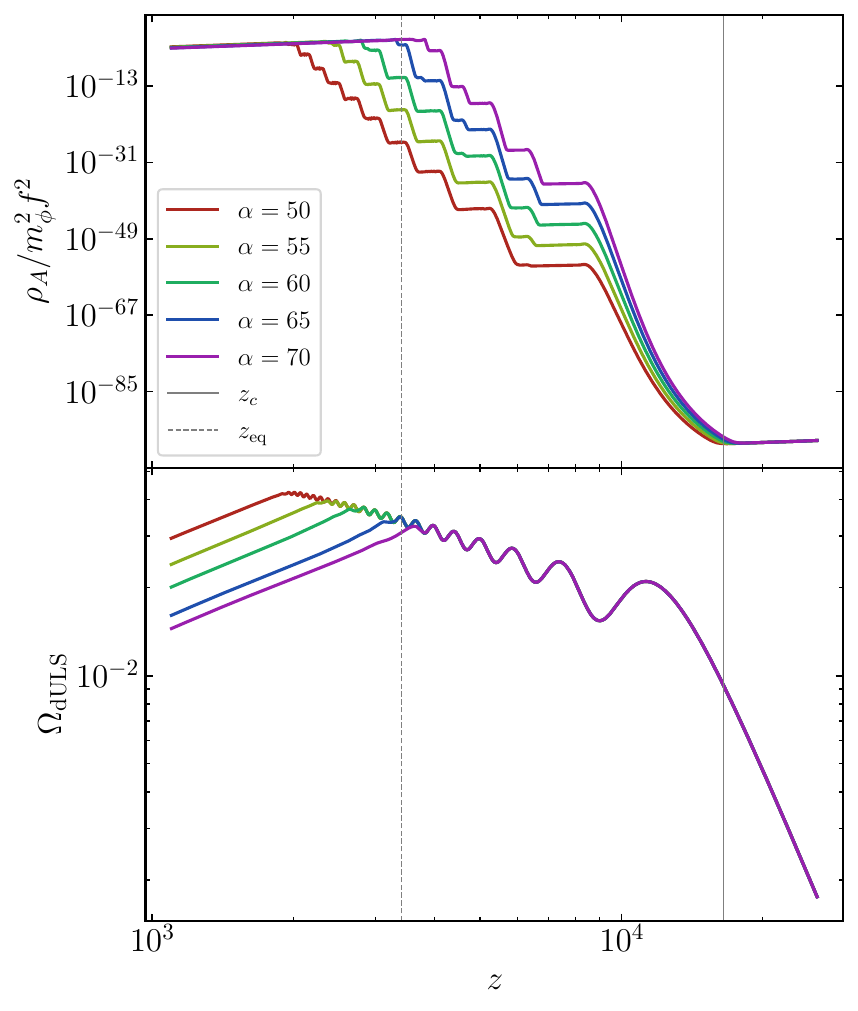}
    \caption{
        Total energy in the gauge fields (top) and the fractional energy in the dULS sector
        (bottom) as a function of redshift for various couplings $\alpha$ indicated in the legend.
        All simulations fix $m_\phi = 10^{-26} \, \mathrm{eV}$,
        $f = 1.5 \times 10^{17} \, \mathrm{GeV}$, and $\theta = 2$.
        Dashed and solid vertical lines indicate, respectively, matter-radiation equality and the
        favored redshift of the transition from dark energy to matter from the analysis of
        Ref.~\cite{Gonzalez:2020fdy}.
    }\label{fig:omega-duls}
\end{figure}

We first verify that the dynamics of the dULS sector qualitatively reproduce the effective fluid
description employed in Ref.~\cite{Gonzalez:2020fdy} (which we evaluate in more detail
in~\cref{app:compare-two-fluid}).
In \cref{fig:omega-duls} we display the energy in the gauge fields $\rho_A$ and the fractional
energy in the dULS sector, $\Omega_\mathrm{dULS} = (\rho_A + \rho_\phi) / \rho$, as a function of
redshift.
We vary the coupling $\alpha$ from 50 to 70, spanning values large enough for resonance and GW
production to terminate before recombination while small enough to be reliably resolved by our grid.

\begin{figure}[t]
    \centering
    \includegraphics[width=\columnwidth]{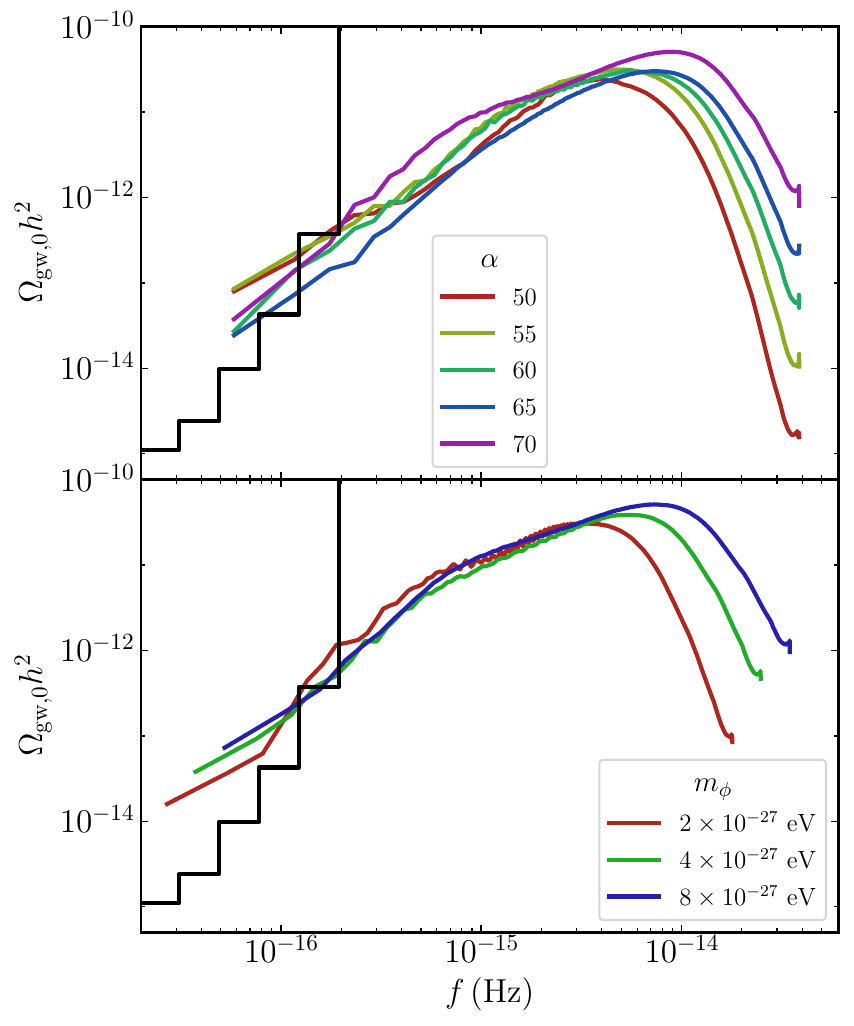}
    \caption{
        Present-day gravitational wave spectrum emitted by recombination (i.e., evaluated at $z
        \approx 1100$), fixing $m_\phi = 10^{-26} \, \mathrm{eV}$ and varying $\alpha$ (top) and
        fixing $\alpha = 70$ and varying $m_\phi$ (bottom).
        All simulations set $f = 1.5 \times 10^{17} \, \mathrm{GeV}$ and $\theta = 2$.
        In black is the upper bound as constrained by the CMB, computed in
        Ref.~\cite{Clarke:2020bil}.
    }\label{fig:vary-af-and-mphi}
\end{figure}

\Cref{fig:vary-af-and-mphi} depicts the gravitational wave signal resulting from the resonant
production of dark photons, evaluated at $z = 1100$.
We compare the signal to the constraints from a recent analysis of current data in
Ref.~\cite{Clarke:2020bil}.\footnote{
    Note that the constraints of Ref.~\cite{Clarke:2020bil} are not directly applicable to this
    scenario, since they are computed from adiabatic initial conditions---constant initial GW
    amplitude on superhorizon scales.
    By contrast, the GWs here are actively sourced, analogous to those in defect scenarios, e.g.,
    Ref.~\cite{Seljak:1997ii}.
    However, we expect constraints on active modes to be competitive, if not more severe than those
    on adiabatic modes.
}
While the peak of the fractional gravitational wave energy spectrum (reaching above $10^{-11}$)
resides at larger frequencies $\sim 10^{-14} \, \mathrm{Hz}$, the infrared tail of the spectrum
exceeds constraints at frequencies $\lesssim 10^{-16} \, \mathrm{Hz}$ by roughly an order of
magnitude.
We note that, though we are unable to simulate larger volumes to capture even lower frequencies, we
expect the signal should continue roughly as a power law further into the infrared,
$\Omega_{\rm gw} \propto k/k_\star$, where $k_\star$ is the peak wave number, as suggested by
recent analytic estimates of GW production in similar scenarios~\cite{Salehian:2020dsf}.

The analysis of Ref.~\cite{Gonzalez:2020fdy} determined that the dULS sector should begin to decay
like radiation by a redshift between $\sim 11000$ and $5000$; as a result, if this scenario is to
alleviate the Hubble tension, the associated gravitational wave signal will be produced before
recombination, $z \approx 1400 - 1100$~\cite{Caprini:2018mtu}.
While the couplings we are able to study here only probe transitions to radiation-like behavior at
the later end of this interval, our findings offer no reason to expect the signal from models with
larger couplings to evade CMB constraints.

We now consider the dependence of gravitational wave production on other model parameters.
The scaling of the signal with the axion mass $m_\phi$ is relatively simple. From the transfer
function (derived in~\cref{app:equations}), the present-day frequency scales as
$f \sim k / \sqrt{H \Mpl} \sim \sqrt{m_\phi / \Mpl}$.
While the axion mass scales out of the dynamics, it has an effect on the initial amplitude of gauge
field vacuum fluctuations.
A lower mass sets initial fluctuations with lower amplitude, requiring a longer period of resonance
to fully deplete the axion's energy.
However, this effect is relatively unimportant and is easily compensated for by a slight
increase in the coupling $\alpha$.
At smaller masses, resonance begins later; therefore, marginally larger couplings are required in
order for the process to complete before recombination (so that the signals are detectable).
As can be seen in the bottom panel of \cref{fig:vary-af-and-mphi}, this condition is easily met with
$\alpha = 70$ and a variety of masses between $10^{-27}$ and $10^{-26} \, \mathrm{eV}$.
Furthermore, the shape and amplitude of the GW signal itself is qualitatively independent of the
axion mass $m_\phi$.

The value of the axion decay constant $f$ is set by the requirement that the dULS sector
comprises a fraction of the Universe's energy between $3-4\%$, leaving little room for variation.
Namely, at the onset of oscillations, the fractional energy in the axion scales as
$\rho_\phi / \rho \sim m_\phi^2 (\theta f)^2 / H^2 \sim (\theta f)^2$.
In turn, the amplitude of the resulting gravitational wave signal is directly proportional to the
square of the fraction of the Universe's energy residing in its source~\cite{Giblin:2014gra}.
Since the signals we find here exceed current constraints by an order of magnitude, we
do not expect the constraining power of the GW signal to be sensitive to any uncertainty in the
best-fit $\rho_\phi / \rho$.

Since $\theta$ sets the amplitude of axion oscillations (and so
$\left\langle \phi' \right\rangle$), its effect on the dynamics is, to linear order, degenerate
with the coupling.
However, nonlinear effects (e.g., rescattering of power to higher momenta) become more important
with larger couplings $\alpha$, and so our choice of $\theta = 2$ allows reliable simulations with
smaller couplings that still transition the dULS sector to a radiation-like state on the required
time scales.

\begin{figure}[t]
    \centering
    \includegraphics[width=\columnwidth]{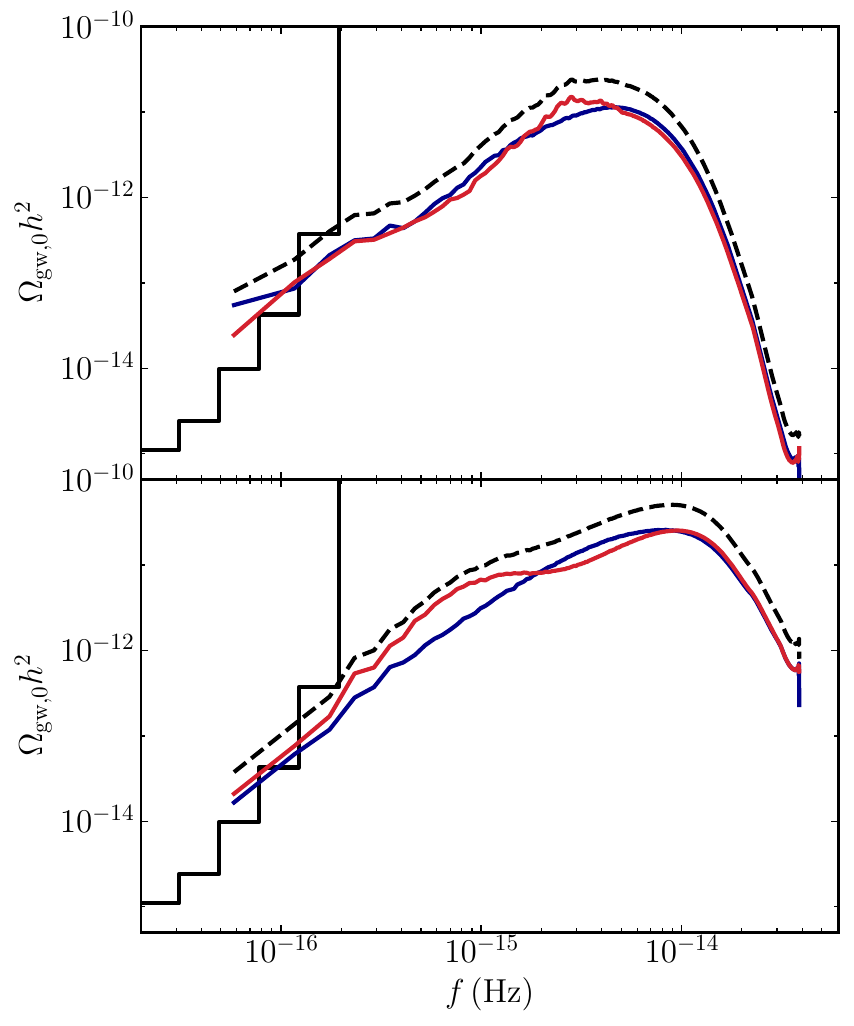}
    \caption{
        Polarization components of the present-day gravitational wave spectrum emitted by
        recombination (i.e., evaluated at $z \approx 1100$) for $\alpha = 50$ (top) and $70$
        (bottom).
        The plus and minus polarizations are in blue and red, respectively, while the total signal
        is portrayed in dashed black.
        Both panels fix $m_\phi = 10^{-26} \, \mathrm{eV}$,
        $f = 1.5 \times 10^{17} \, \mathrm{GeV}$, and $\theta = 2$.
        In solid black is the upper bound as constrained by the CMB, computed in
        Ref.~\cite{Clarke:2020bil}.
    }\label{fig:gw-pol}
\end{figure}

As a final investigation, we study whether the gravitational wave signal is significantly polarized.
The same axial coupling of gauge fields to the inflaton generates a helical gravitational wave
background during inflation~\cite{Cook:2011hg,Barnaby:2011qe,Anber:2012du,Domcke:2016bkh,
Bartolo:2016ami}, and can also imprint on the spectrum of gravitational waves
produced during preheating~\cite{Adshead:2018doq}.
However, in the former case, the sign of the axion's velocity is fixed during inflation.
Preheating via the axial coupling can also complete within one (or even half an) oscillation of the
inflaton~\cite{Adshead:2015pva,Adshead:2019lbr,Adshead:2019igv}, but nonlinear effects can result in
a gravitational wave signal dominated by different helicities at different scales.
The results presented in \cref{fig:gw-pol} follow in spirit.
For the lowest coupling we consider, $\alpha = 50$, the axion oscillates numerous times before gauge
field production terminates, emitting an essentially unpolarized gravitational wave background.
For the largest coupling, $\alpha = 70$, the spectrum is moderately polarized at large scales,
consistent with a more substantial enhancement of one polarization before the axion first crosses
zero.
While the signal at lower frequencies arises predominantly from the initial phase of helical
tachyonic resonance, higher frequencies are sourced by nonlinear mode interactions which do not
retain the same polarization.
In summary, it is difficult to evaluate whether the gravitational wave background is sufficiently
polarized on CMB scales to provide a unique signature of this model; doing so likely requires
thorough study via numerical simulations in a model-dependent way.
As an aside, we note that these findings are applicable to models of dark matter as massive dark
photons, which are produced via the same resonant instability considered
here~\cite{Agrawal:2018vin,Machado:2018nqk,Machado:2019xuc,Bastero-Gil:2018uel,Co:2018lka,
Dror:2018pdh}.

\section{Conclusions}

The rapid production of inhomogeneities from resonant particle production
can induce a significant GW background.
The amplitude of the GW signal is largest (and so offers the most constraining potential) when the
GW source comprises a significant fraction of the Universe's energy budget and
occurs close to the horizon scale at the time of emission~\cite{Giblin:2014gra}.
The model considered here, which exhibits a tachyonic instability via an axion coupled to dark
photons, is especially efficient, as has been shown in the context of preheating after
inflation~\cite{Adshead:2019lbr,Adshead:2019igv} in which case up to the entire energy budget of
the Universe may source gravitational waves.
New physics which relies on the same mechanism later in cosmological history is subject to
constraints from direct probes of stochastic backgrounds of gravitational
waves~\cite{Lasky:2015lej,Caprini:2018mtu}.
Models of early dark energy proposed to alleviate the Hubble tension are a prime example, as their
success hinges on the new sector making up a substantial [$\mathcal{O}(1\%)$] fraction of the
Universe's energy.
Furthermore, the source of early dark energy must soon decay one way or another before
recombination, pinning the relevant length scales to those probed by the cosmic microwave
background.

In this work, we demonstrated that the decaying ultralight scalar model, as motivated in
Ref.~\cite{Gonzalez:2020fdy}, produces a background of gravitational waves with a peak spectral
energy fraction exceeding $\mathcal{O}(10^{-11})$ at its peak, with a power-law tail extending into
the region that is already constrained by the CMB~\cite{Clarke:2020bil}.
While we have not studied the entire available parameter space, we showed that the requirements for
the model to successfully alleviate the Hubble tension generally coincide with those for its
gravitational wave signature to be constrained by the CMB.
For the parameter space we considered, we found that the signal exceeds constraints (on adiabatic
modes) by an order of magnitude.
Because GWs are actively sourced in this scenario, these constraints are not directly applicable,
and we leave a detailed computation of the CMB signatures of these models to future work.
However, we expect the importance of this differing time dependence to be suppressed: in terms of
line-of-sight solutions, the contribution of tensors to the CMB is weighted by the visibility
function, which is sharply peaked at the time of recombination~\cite{Zaldarriaga:1996xe}.
In addition, recent work has proposed spectral distortions as a probe of gravitational waves at
higher frequencies than those directly probed by the CMB~\cite{Kite:2020uix}, which would be
sensitive to the peak of the signals presented here.

Finally, we point out that resonant particle production is not a unique feature of this model.
The original single-field models of EDE exhibit similar parametric instabilities which may also emit
significant GW backgrounds~\cite{Smith:2019ihp}.
More broadly, our findings suggest that stochastic backgrounds of gravitational waves could provide
an orthogonal probe with which to constrain models of early dark energy.


\begin{acknowledgements}
We gratefully acknowledge Tom Clarke, Ed Copeland, and Adam Moss for sharing the data from
Ref.~\cite{Clarke:2020bil}, and we thank Tristan Smith for useful discussions.
The work of P.A.\ was supported in part by NASA Astrophysics Theory Grant NNX17AG48G.
J.T.G.\ is supported by the National Science Foundation Grant No. PHY-1719652.
Z.J.W.\ is supported in part by the United States Department of Energy Computational Science
Graduate Fellowship, provided under Grant No. DE-FG02-97ER25308.
This work used the Extreme Science and Engineering Discovery Environment (XSEDE)~\cite{xsede}, which
is supported by National Science Foundation grant number ACI-1548562; simulations were run on the
Comet cluster at the San Diego Supercomputer Center through allocation TG-PHY180049.
This work made use of the Illinois Campus Cluster, a computing resource that is operated by the
Illinois Campus Cluster Program (ICCP) in conjunction with the National Center for Supercomputing
Applications (NCSA) and which is supported by funds from the University of Illinois at
Urbana-Champaign.

Simulations in this work were implemented with \textsf{pystella}, which is available at
\href{https://github.com/zachjweiner/pystella}{github.com/zachjweiner/pystella} and makes use of
the Python packages \textsf{PyOpenCL}~\cite{kloeckner_pycuda_2012},
\textsf{Loo.py}~\cite{kloeckner_loopy_2014}, \textsf{mpi4py}~\cite{DALCIN2008655,DALCIN20051108},
\textsf{mpi4py-fft}~\cite{jpdc_fft}, \textsf{NumPy}~\cite{numpy}, and \textsf{SciPy}~\cite{scipy}.
\end{acknowledgements}




\appendix


\section{Equations and numerical implementation}\label{app:equations}

The dynamical system, our conventions, and our numerical implementation match those of
Refs.~\cite{Adshead:2019lbr,Adshead:2019igv} (see the appendices of Ref.~\cite{Adshead:2019lbr} for
full details), with relatively minimal differences we describe here.

The energy density and pressure which source FLRW expansion comprise contributions from the standard
background $\Lambda$CDM model, the axion, and the gauge fields.
The spatially-averaged energy density and pressure are
\begin{align}
    \label{eqn:phi-rho}
    \rho_\phi(\tau)
    &\equiv \left\langle
            \frac{{\phi'}^2}{2 a^2}
            + \frac{(\partial_i \phi)^2}{2 a^2}+ V(\phi)
        \right\rangle, \\
    \label{eqn:phi-pressure}
    p_\phi(\tau)
    &\equiv \left\langle
            \frac{{\phi'}^2}{2 a^2}
            - \frac{(\partial_i \phi)^2}{6 a^2}
            - V(\phi)
        \right\rangle
\end{align}
for the axion and
\begin{align}
    \label{eqn:gauge-rho}
    \rho_A(\tau)
    &\equiv  \frac{1}{2} \left\langle {\mathbf{E}}^2 + {\mathbf{B}}^2 \right\rangle, \\
    \label{eqn:gauge-pressure}
    p_A(\tau)
    &\equiv \frac{1}{6} \left\langle {\mathbf{E}}^2 + {\mathbf{B}}^2 \right\rangle
\end{align}
for the gauge fields.
Above we defined the electric and magnetic fields
\begin{align}
    \label{eq:electric_magnetic}
    E_{i}
    &= \frac{1}{a^2} \left( {A_{i}}' - \partial_{i} A_0 \right), \quad
    B_{i}
    = \frac{1}{a^2} \epsilon_{i j k} \partial_{j} A_{k}.
\end{align}
In our simulations, we consistently include the contributions of
\cref{eqn:phi-rho,eqn:phi-pressure,eqn:gauge-rho,eqn:gauge-pressure} to the expansion rate alongside
the background $\Lambda$CDM model.

The axion--dark-photon coupling violates parity and so the resonantly produced gauge bosons are
helical (at least initially), which may in principle imprint on the resulting gravitational waves.
For details on the decomposition of gravitational waves into the polarization basis, see
Ref.~\cite{Adshead:2018doq}.

The present-day spectrum $\Omega_\mathrm{gw}(a_0)$ is related to that at the time of emission,
$\Omega_\mathrm{gw}(a_\mathrm{e})$, via
\begin{align}\label{eqn:omega-gw-e-0-general}
    \frac{\Omega_\mathrm{gw}(a_0)}{\Omega_\mathrm{gw}(a_\mathrm{e})}
    = \left( \frac{a_\mathrm{e}}{a_0} \right)^4
        \frac{\rho(a_\mathrm{e})}{\rho(a_0)},
\end{align}
since gravitational waves redshift like radiation.
In turn, the present-day frequency of observation is determined from the physical momentum
$k_\mathrm{phys} = k / a_0$ via
\begin{align}\label{eqn:gw-frequency-general}
    f
    = \frac{k_\mathrm{phys}}{2 \pi}
    = \frac{1}{2 \pi} \frac{k}{a_\mathrm{e}} \frac{a_\mathrm{e}}{a_0}.
\end{align}
The entropy of the Standard Model (SM) provides a conserved quantity by which we may compute
$a_e / a_0$ at any time after the SM thermalizes.
The entropy density is $s = 2 \pi^2 g_{\star S}(T) T^3 / 45$, where $T$ is the SM temperature and
$g_{\star S}$ the number of effective degrees of freedom in entropy.
Combining with $\rho_\mathrm{rad} = \pi^2 g_\star(T) T^4 / 30$ (where $g_\star$ is the
effective number of degrees of freedom in energy density) allows one to express $a_e / a_0$ in terms
of the energy density in radiation, $g_\star$, and $g_{\star S}$.
Approximating $g_{\star S} = g_\star$ allows us to evaluate \cref{eqn:omega-gw-e-0-general} at any
point after thermalization as
\begin{align}
    \Omega_\mathrm{gw}(a_0) h^2
    &= \frac{\Omega_{\mathrm{rad}}(a_0) h^2}{\Omega_{\mathrm{rad}}(a_\mathrm{e})}
        \left( \frac{g_{\star}(a_0)}{g_{\star}(a_\mathrm{e})} \right)^{1/3}
        \Omega_\mathrm{gw}(a_\mathrm{e}).
    \label{eqn:gw-amplitude-transfer-function-after-reheating}
\end{align}
Likewise, \cref{eqn:gw-frequency-general} may be expressed as
\begin{align}
    f
    = \frac{k / 2 \pi a_\mathrm{e}}{\sqrt{H(a_\mathrm{e}) \Mpl}}
        \left(
            \frac{\Omega_{\mathrm{rad}}(a_0)}{\Omega_{\mathrm{rad}}(a_\mathrm{e})}
            H_0^2 \Mpl^2
        \right)^{1/4}
        \left( \frac{g_{\star}(a_0)}{g_{\star}(a_\mathrm{e})} \right)^{1/12}.
\end{align}

While in general $g_{\star}$ may be evaluated using, e.g., tabulated values from
Ref.~\cite{Husdal:2016haj}, the effective number of relativistic degrees of freedom is fixed to
$3.36$ since well before the time of our simulations.
The present radiation fraction $\Omega_{\mathrm{rad}}(a_0) h^2 \approx 4.2 \times 10^{-5}$, and the
radiation fraction at the time of emission is calculated during the simulation.
Plugging in $H_0 = h \cdot 3.2 \times 10^{-18} \, \mathrm{Hz}$ and
$\Mpl = 3.7 \times 10^{42} \, \mathrm{Hz}$,
\begin{align}
    f
    = \frac{k / a_\mathrm{e}}{\sqrt{H(a_\mathrm{e}) \Mpl}}
        \Omega_{\mathrm{rad}}(a_\mathrm{e})^{1/4}
        \times 4.33 \times 10^{10} \, \mathrm{Hz}.
    \label{eqn:gw-transfer-frequency}
\end{align}

To set initial conditions, we begin by solving for the dynamics of the linearized system.
We evolve the Friedmann equations, the homogeneous and linear-order parts of the axion's equation of
motion,
\begin{align}
    \left\langle \phi'' \right\rangle
    &= - 2 \mathcal{H} \left\langle \phi' \right\rangle
        - a^2 \frac{\ud V}{\ud \left\langle \phi \right\rangle}
        - a^2 \frac{\alpha}{4 f} \left\langle F_{\mu \nu} \tilde{F}^{\mu \nu} \right\rangle, \\
    \delta \phi''(k)
    &= - 2 \mathcal{H} \delta \phi'(k)
        - \left(
            k^2
            + a^2 \frac{\ud^2 V}{\ud \left\langle \phi \right\rangle^2}
        \right)
        \delta \phi'(k),
    \label{eqn:delta-phi-linear-eom}
\end{align}
and the linearized equation of motion for the gauge field polarizations,
\begin{align}
    {A_\pm}''(\mathbf{k})
    &= - k \left( k \mp \frac{\alpha}{f} \left\langle \phi' \right\rangle \right) A_\pm(\mathbf{k}).
    \label{eqn:linear-eom-gauge-polarization}
\end{align}
While solving the linearized system, we compute the gauge field contributions to the background
dynamics by integrating over the evolved gauge field modes via numerical quadrature
(as employed in Ref.~\cite{Adshead:2019lbr}).
We neglect the backreaction of the gauge fields onto the axion fluctuations in
\cref{eqn:delta-phi-linear-eom}, which is negligible even well after we initialize the full
lattice simulation.

We begin the linear evolution when $H \gg m_\phi$ (so that all modes are far outside the horizon)
and set $\left\langle \phi \right\rangle = \theta f$ and $\left\langle \phi' \right\rangle = 0$.
The gauge fields are initialized in the Bunch-Davies vacuum,
\begin{align}
	\left\langle \left\vert A_\pm(\mathbf{k}) \right\vert^2 \right\rangle
    &= \frac{1}{2 \sqrt{k}}, \\
	\left\langle \left\vert A_\pm'(\mathbf{k}) \right\vert^2 \right\rangle
    &= \frac{\sqrt{k}}{2}.
\end{align}
The axion acquires a nearly scale-invariant spectrum of fluctuations during inflation,
\begin{align}
	\left\langle \left\vert \phi(\mathbf{k}) \right\vert^2 \right\rangle
    &= \frac{1}{k^3} \frac{H_\mathrm{inf}^2}{2 \pi^2},
\end{align}
and $\delta \phi' = 0$.
As a benchmark value, we take the inflationary Hubble scale
$H_\mathrm{inf} = 2 \times 10^{9} \, \mathrm{GeV}$, but our main results are insensitive to whether
the axion is even initialized with any fluctuations.

We evolve the linear system until $H = m_\phi$, at which point we initialize the lattice simulation.
The subsequent dynamics are identical to when the lattice simulation is initialized when
$H = 10 m_\phi$, and are relatively unchanged even if initialized at $H = m_\phi / 10$.
Using the background values and power spectra obtained at this point ($H = m_\phi$), we initialize
the lattice simulation with the procedure described in Ref.~\cite{Adshead:2019lbr}.
Our simulations use a grid with $N^3 = 768^3$ points and conformal box length $L = 10 / m_\phi$,
with a time step $\Delta \tau = \Delta x / 10 = L / 10 N$.
Though our computational resources only allowed for a single simulation (i.e., random realization of
the initial conditions) per set of model parameters, we do not expect an ensemble average to
exhibit significant variance in, e.g., the gravitational wave spectrum.


\section{Convergence tests}\label{sec:convergence}

\begin{figure}[t!]
    \centering
    \includegraphics[width=\columnwidth]{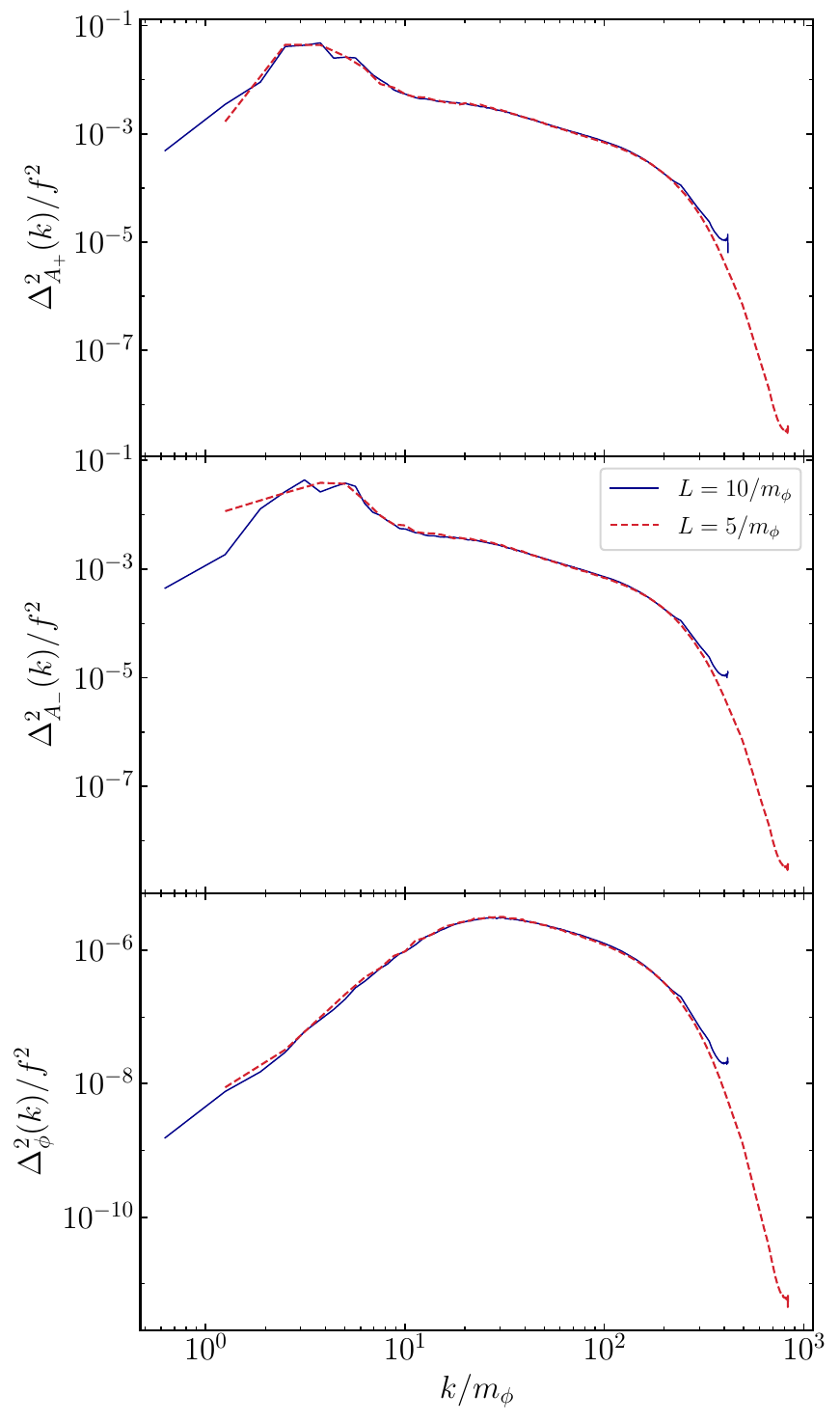}
    \caption{
        Dimensionless power spectra of $A_+$, $A_-$ and $\phi$ (from top to bottom), evaluated
        at the end of the simulation ($z = 1100$), defined by \cref{eqn:dimless-power-spectrum}.
        All simulations fix $\alpha = 70$, $m_\phi = 10^{-26} \, \mathrm{eV}$,
        $f = 1.5 \times 10^{17} \, \mathrm{GeV}$, and $\theta = 2$.
        We compare simulations with box lengths $L = 10 / m_\phi$ in solid, thin blue and
        $L = 5 / m_\phi$ in dashed red.
    }\label{fig:convergence}
\end{figure}

The simulations we present in the main text require fairly large volumes to capture the IR tail of
the gravitational wave spectrum which resides at CMB frequencies, while also sufficient resolution
to capture both the initial tachyonic resonance band and the subsequent power transfer to higher
momenta via nonlinear effects.
As $N^3 = 768^3$ represents the largest grid size possible with our current resources, in lieu of an
ideal convergence test (fixing the box length $L$ while increasing $N$) we compare results for two
box lengths, both with $N = 768$.
We present this comparison for $L = 10 / m_\phi$ (as used for main results) and $L = 5 /
m_\phi$ in \cref{fig:convergence}, for $\alpha = 70$ (the largest coupling for which we present
results, which should require the most resolution).
Specifically, \cref{fig:convergence} depicts for each field the final dimensionless power spectra,
i.e., for a field $f$,
\begin{align}\label{eqn:dimless-power-spectrum}
	\Delta_f(k)^2
	= \frac{1}{2\pi^2} \frac{1}{V} \int \frac{\ud \Omega}{4\pi} \, k^3 \abs{ f(\mathbf{k}) }^2.
\end{align}
The results show excellent consistency, with expected discrepancy at the largest scales in the
simulation (due to statistical variance) and negligible differences at the Nyquist frequency.


\section{Comparing to the two-fluid description}\label{app:compare-two-fluid}

\begin{figure}[th]
    \centering
    \includegraphics[width=\columnwidth]{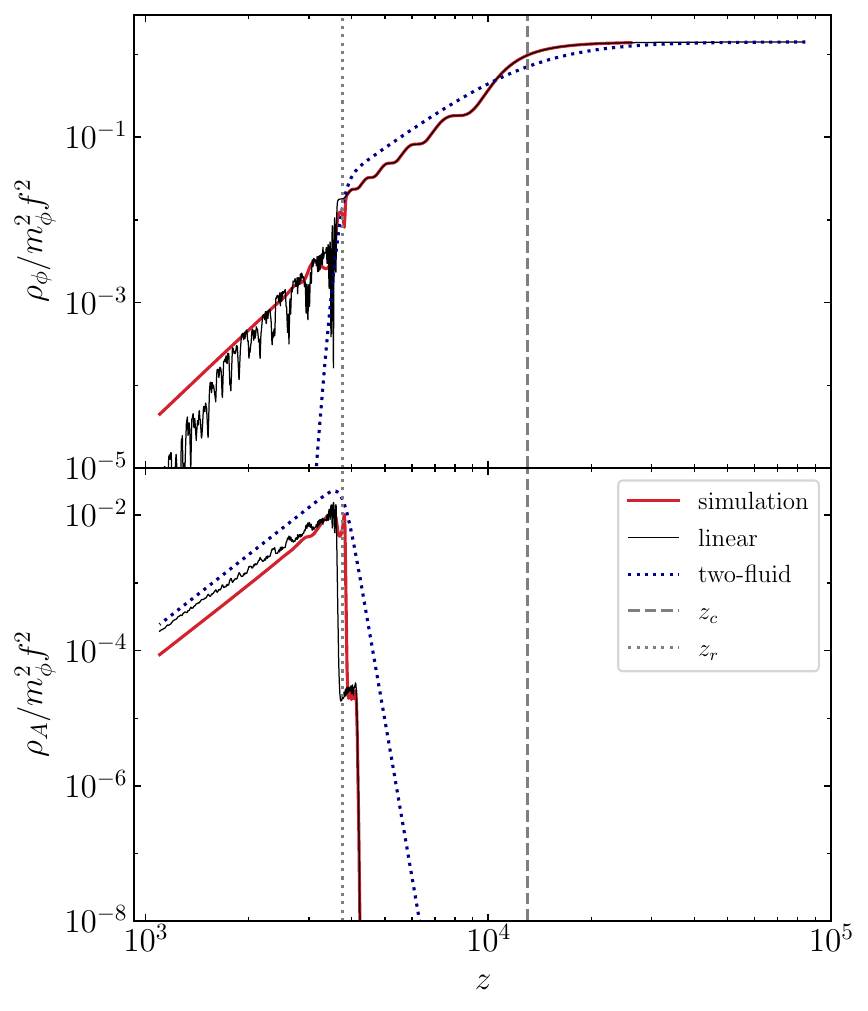}
    \caption{
        Comparing numerical solutions for the full nonlinear system of equations (red), the
        linearized system (thin black), and the effective two-fluid description of
        Ref.~\cite{Gonzalez:2020fdy} (dotted blue).
        Vertical lines denote the redshift of oscillation $z_c = a_0 / a_c - 1$ (dashed grey)
        and that of the transition to radiation domination $z_r = a_0 / a_r - 1$ (dotted grey).
        For the fluid model, we choose $a_c = 6.5$, $a_r = 3.5 a_c$ to be consistent with the
        dynamics of the simulation and linear solution, which take
        $m_\phi = 10^{-26} \, \mathrm{eV}$, $\theta = 2$, $\alpha = 70$, and
        $f = 1.5 \times 10^{17} \, \mathrm{GeV}$.
    }\label{fig:sim-linear-two-fluid-compare}
\end{figure}

We now compare the results of the full numerical simulation, the numerical solution to the
linearized system, and the two-fluid model detailed in Ref.~\cite{Gonzalez:2020fdy}.
The latter is described by the effective fluid equations
\begin{align}
    \rho_\phi'
    + 3 a H (1 + w_\phi) \rho_\phi
    &= - a \Gamma(\tau) \rho_\phi, \\
    \rho_A'
    + 4 a H \rho_A
    &= a \Gamma(\tau) \rho_\phi,
\end{align}
where the axion equation of state is
\begin{align}
    w_\phi(a)
    = -1 + \frac{1}{1 + (a_c / a(\tau))^3}
\end{align}
and the decay rate is parameterized by
\begin{align}
    \Gamma(\tau)
    = \frac{m_\phi}{1 + (a_r / a(\tau))^p}
\end{align}
with $p = 30$.
In this effective description, the scalar field begins to oscillate at $a_c$, while the dULS sector
transitions to radiation domination at $a_r$.
We display the evolution of $\rho_\phi$ and $\rho_A$ in \cref{fig:sim-linear-two-fluid-compare}.
Qualitatively, the two-fluid model reproduces the energy transfer, but cannot account for the
remnant energy remaining in the axion after resonance.
This remaining energy is due to nonlinear effects preventing the axion condensate from being totally
depleted as well as backreaction effects producing axion particles.
In addition, \cref{fig:sim-linear-two-fluid-compare} demonstrates the validity of the solution to
the linearized system of equations in the early stages of resonance while also exhibiting the
importance of nonlinear effects as resonance terminates.

\bibliography{references}

\end{document}